\documentclass{article}
\usepackage{amsmath}

\usepackage{PRIMEarxiv}
\usepackage[utf8]{inputenc} % allow utf-8 input
\usepackage[T1]{fontenc}    % use 8-bit T1 fonts
\usepackage{hyperref}       % hyperlinks
\usepackage{url}            % simple URL typesetting
\usepackage{booktabs}       % professional-quality tables
\usepackage{amsfonts}       % blackboard math symbols
\usepackage{nicefrac}       % compact symbols for 1/2, etc.
\usepackage{soul}
\usepackage{microtype}      % microtypography
\usepackage{lipsum}
\usepackage{fancyhdr}       % header
\usepackage{graphicx}       % graphics
\graphicspath{{media/}}     % organize your images and other figures under media/ folder
\usepackage{subfigure}
\usepackage{graphicx} %插入图片的宏包
\usepackage{tabularx} %by ZW
\usepackage{amsmath} % 需要引入amsmath宏包
\title{Voltage Regulation in Polymer Electrolyte Fuel Cell Systems Using Gaussian Process Model Predictive Control}

\author{
\textbf{Xiufei Li} $^{1\ast}$ \quad
\textbf{Miao Zhang} $^{2\ast}$  \quad  
\textbf{Yuanxin Qi}  $^{1}$ \quad 
\textbf{Miao Yang}  $^{3\ast\ast}$ \\
$^{1}$ Lund University, Department of Energy Sciences, Lund, 22100, Sweden  \\
$^{2}$ Shenzhen International Graduate School,Tsinghua University, Shenzhen, 518055, China\\
$^{3}$ City University of Hong Kong, Department of Mechanical Engineering, Hong Kong, 999077, China\\
$^{\ast\ast}$ Corresponding author: Miao Yang, yang11415@163.com\\
$^{\ast}$ First Author and Second Author contribute equally to this work.}

\begin{document}
\maketitle
\begin{abstract}
This study introduces a novel approach utilizing Gaussian process model predictive control (MPC) to stabilize the output voltage of a polymer electrolyte fuel cell (PEFC) system by simultaneously regulating hydrogen and airflow rates. Two Gaussian process models are developed to capture PEFC dynamics, taking into account constraints including hydrogen pressure and input change rates, thereby aiding in mitigating errors inherent to PEFC predictive control. The dynamic performance of the physical model and Gaussian process MPC in constraint handling and system inputs is compared and analyzed. Simulation outcomes demonstrate that the proposed Gaussian process MPC effectively maintains the voltage at the target 48 V while adhering to safety constraints, even amidst workload disturbances ranging from 110-120 A. In comparison to traditional MPC using detailed system models, Gaussian process MPC exhibits a 43\% higher overshoot and 25\% slower response time. Nonetheless, it offers the advantage of not requiring the underlying true system model and needing less system information.

\end{abstract}

% % keywords can be removed
% \keywords{Polymer electrolyte fuel cell \sep Gaussian process \sep Model predictive control (MPC) \sep Constraints}

\section{Introduction}
\label{Introduction}

Due to the adverse environmental consequences associated with traditional fossil fuels and their contribution to global warming, PEFCs have emerged as highly promising power sources owing to their utilization of renewable energy sources, high energy efficiency, and low operating temperatures \cite{baroutaji2021advancements,zhang2020thermal,dong2021honeycomb,xu2021analysis}. While PEFCs are increasingly favored across stationary, portable, and transportation applications, their commercialization remains hampered by various technical challenges, with ensuring reliable operation standing out as a major hurdle \cite{liang2021electrocatalytic,cao2020efficient,parbey2020electrospun}. Control algorithms play a pivotal role in enhancing the reliability of PEFC systems. However, the complexity of PEFC systems, characterized by numerous input parameters and safety constraints, necessitates the adoption of more sophisticated control methodologies \cite{silaa2020design}.

MPC controllers distinguish themselves in PEFC systems' control applications due to their robust handling of multiple inputs and constraints, making them widely utilized across various PEFC system applications.\cite{quan2021feedback} introduced a multi-input multi-output (MIMO) MPC to regulate hydrogen excess ratio and balance electrode pressures. \cite{hahn2021adaptive} proposed an MPC-based operation strategy for controlling an automotive fuel cell air system. Their study demonstrated that the MPC approach could potentially reduce hydrogen consumption by 3\% while decreasing the risk of harmful operating conditions compared to a validated map-based operation strategy. \cite{long2015current} devised a master-slave MPC method for current-sharing control in systems, demonstrating the feasibility and effectiveness of the proposed MPC-based current-sharing strategy. Recent these studies have showcased various approaches to MPC applications, highlighting its efficacy and promise as a control method.

The MPC models employed in the aforementioned research, however, primarily rely on physical models. The modeling process entails numerous partial differential equations, and simplifying them necessitates expert knowledge and is time-intensive. Conversely, data-based modeling leverages observed data rather than intrinsic physical laws to derive the system model. Data-based models have garnered increasing attention due to their robust representation capabilities, flexibility, and ease of construction. The MPC control approaches derived from these data-driven models hold appeal for fuel cell applications, given the nonlinearity and complexity inherent in fuel cell systems. \cite{MEHRPOOYA20188} trained a two-hidden-layer neural network to predict fuel cell steady-state performance using experimentally collected data encompassing various inlet humidity, temperature, and oxygen and hydrogen flow rates. \cite{HAN201610202} employed both neural networks and support vector machines to characterize fuel cell polarization curves and compared their performances. \cite{HAN201610202} utilized neural networks for fuel cell dynamics prediction, trained on data collected from MATLAB simulations. While one frequently employed data-based method is the Gaussian process (GP) \cite{PILLONETTO2014657}. Unlike optimizing parameters of selected functions to fit data, GP searches for relationships between measured data. \cite{He2016GP} employed a nonparametric Gaussian process regression model to capture the nonlinear relationship between operating conditions and output voltage in microbial fuel cells. They also proposed a simple online learning strategy for recursively updating model hyperparameters. \cite{ZHU201863} applied Gaussian process state-space models to analyze fuel cell degradation and incorporated prediction confidence intervals to enhance inference accuracy. \cite{Zhang2020Fueladaptive} constructed a Gaussian process regression model to predict methane conversion rates in solid oxide fuel cells.

However, a comprehensive model was lacking to consider the constraints associated with control variables and hydrogen pressure. This paper addresses this gap by imposing limits on both control variables and hydrogen pressure to ensure safe operation. Specifically, when utilized as a power generation source, a fuel-cell system must deliver a stable and consistent voltage \cite{li2021adaptive}. In our previous study \cite{li2021multi}, a novel MPC approach was devised to regulate the output voltage of the PEFC system to a steady state by simultaneously adjusting its input hydrogen and air volumetric flow rates. The simulation results demonstrate the MPC controller's notable effectiveness in stabilizing the fuel cell system's voltage, exhibiting significantly reduced overshoot and faster response times.  Furthermore, a Gaussian process MPC controller is developed for voltage control of the PEFC system. Gaussian processes are employed to model both the voltage and hydrogen pressure dynamics. The capability of the Gaussian process MPC to manage the fuel cell system under constraints is thoroughly investigated, and a performance comparison between MPC based on physical models and Gaussian process MPC is presented.

The remaining sections of the paper are structured as follows: Section 2 introduces the Gaussian process modeling process; Section 3 elaborates on the development of the GP-MPC algorithm; Section 4 describes the simulation setup; Section 5 presents the performance of the GP-MPC and compares it to MPC based on physical models; Section 6 provides a discussion; and finally, Section 7 concludes the paper.

\section{PEFC System Model Development}
\subsection{Preanalysis of the PEFC model}
The PEFC model developed in this paper is based on previously proposed semiempirical models {\cite{wu2008diagnostic}}, which have the capabilities to illuminate the electrochemical behavior of a FC without offering deep apprehension of the underlying phenomena {\cite{ettihir2014online}}. The PEFC model developed in this work is on the system scale. Therefore, the focus is on the macro performances of the PEFC system, the details of fluid flow in the porous area, the temperature distribution across the cell area, and the reaction distribution on the catalyst layer are not considered. However, the electrochemical reaction determines the transfer of electrons and protons, which will affect the stack temperature and the FC performance. In the temperature range of a PEFC system (223-373 K), we may assume the reactant gases to follow the ideal gas law, e.g. comparing the operating temperature to the critical temperature of each component. The water generated in the cathode may involve two-phase flow pattern, which is beyond the scope of this paper, and the water can be drained from the system, thus not considered.

\subsection{PEFC System Model}
Through a pair of redox reactions, the hydrogen fed PEFC converts the chemical energy from hydrogen and oxygen into electricity with only heat and water as byproducts. Its typical output voltage is usually less than the ideal value because of some losses occurred inside the fuel cells \cite{correa2004electrochemical,zhang2019numerical}. Thus, to get higher voltage, a number of cells are usually combined in series and the net output voltage of a PEFC 
is given as follows \cite{kandidayeni2019benchmark}:
\begin{equation}
 V_{FC}=n_{cell}(E_{nernst}-V_{act}-V_{ohmic}-V_{con})     
\end{equation}
here, {\em V$_{FC}$}, {\em n$_{cell}$}, {\em E$_{nernst}$}, {\em V$_{act}$}, {\em V$_{ohmic}$} and {\em V$_{con}$} denote output voltage of the fuel cell system, cell numbers, reversible voltage, activation voltage drop, ohmic voltage drop and concentration voltage drop.
{\em E$_{nernst}$} is calculated based on the Nernst equation \cite{kandidayeni2019benchmark}:
\begin{equation} \label{eq:E_nernst}
E_{nernst}=1.229-0.85*10^{-3}(T_{stack}-298.15)+4.3085*10^{-5}T_{stack}\left [ln(P_{H_{2}}) +0.5ln(P_{O_{2}})\right]    
\end{equation}
here, {\em T$_{stack}$}, {\em P$_{H_2}$} and {\em P$_{O_2}$} are stack temperature, hydrogen pressure and oxygen pressure.
The activation voltage drop {\em V$_{act}$} occurs due to the activation of the electrodes, and it is defined as\cite{kandidayeni2019benchmark}:
\begin{equation}
V_{act}=-\left [ \xi _{1}+\xi_{2}T_{stack}+\xi _{3}T_{stack}ln(C_{O_{2}})+\xi _{4}T_{stack}ln\left ( I \right )\right ]  
\end{equation}
\begin{equation}
C_{O_{2}}=\frac{P_{O_{2}}}{5.08*10^{6}*e(\frac{-498}{T_{stack}})}
\end{equation}
here, {\em $\xi$} is the semi-empirical coefficient \cite{mann2000development}, {\em C$_{O_2}$} is the oxygen concentration, and {\em I} is the current. The ohmic voltage drop {\em V$_{ohmic}$} comes from the resistance to the electrons transfer and protons transfer. It is given as \cite{kandidayeni2019benchmark}:
\begin{equation}
 V_{ohmic}=I(R_{m}+R_{C})     
\end{equation}
\begin{equation}
 R_{m}=\frac{\rho _{m}l}{A}     
\end{equation}
\begin{equation}
 \rho _{m}=\frac{181.6\left [ 1+0.03(i)+0.062(T_{stack}/303)^{2}(i)^{2.5}\right ]}{(\lambda -0.643-3*i)exp(4.18(\frac{T_{stack}-303}{T_{stack}}))}     
\end{equation}
Here, {\em $R_{m}$}, {\em $R_{C}$}, {\em $\rho _{m}$}, $l$, $A$, $i$, {\em $\lambda$} represent membrane resistance, equivalent
contact resistance to electron conduction, membrane resistivity, membrane thickness, membrane active area, actual current density and adjustable parameter dependent on membrane water content of the membrane. The concentration voltage drop {\em V$_{con}$} is because of the mass transfer which reduces the reactants pressure, and it is determined as \cite{kandidayeni2019benchmark}:
\begin{equation}
V_{con}=-\beta ln(1-i/J_{max})
\end{equation}
Here, {\em $\beta$} is a parametric coefficient related to the fuel cell operating condition, {\em $J_{max}$} denotes the maximum current density.
The dynamic behavior of a PEFC is largely affected by a 'charge double layer' phenomenon. The charge layer on the interface electrode/electrolyte acts as an electrical capacitor. There is always a delay for the charge layer to follow the voltage changes. This delay only affects the activation and concentration voltage drop, which can be described as the following equations\cite{correa2004electrochemical}: 
\begin{equation}
\frac{dV_{a}}{dt}=\frac{I}{C}-\frac{V_{a}}{R_{a}C}
\end{equation}
\begin{equation}
R_{a}=\frac{V_{act}+V_{con}}{I}
\end{equation}
here, {\em $C$} and {\em $R_{a}$} denote the equivalent capacitance of the system and the equivalent resistance. Thus, the output voltage of the PEFC can be rewritten as:
\begin{equation}\label{eq:V_FC rewritten}
V_{FC}=n_{cell}(E_{nernst}-V_{a}-V_{ohmic}) 
\end{equation}

\subsection{Gaussian process}

A Gaussian process is a probabilistic, non-parametric black-box model that is defined as a collection of random variables. Any finite number of variables from this collection have a joint Gaussian distribution \cite{GaussianProcessesforMachineLearning2006}. A Gaussian process $\mathcal{GP}$ can be fully characterized by a mean function $\bar{f}(\boldsymbol{x})$ and covariance function
$k(\boldsymbol{x},\boldsymbol{x}')$,
\begin{equation} \label{eq:GP}
    f(\boldsymbol{x})\sim \mathcal{GP}\left(\bar{f}(\boldsymbol{x}),k(\boldsymbol{x},\boldsymbol{x}')\right),
\end{equation}
where $\boldsymbol{x}$ and $\boldsymbol{x}'$ both are data point with dimension $d$. $k(\boldsymbol{x},\boldsymbol{x}')$ is also called the kernel function.
The mean function $\bar{f}(\boldsymbol{x})$ is assumed to be zero to simplify the analysis.
% Here the $\bar{f}(\boldsymbol{x})$ is taken to zero for simplicity.

The collected data points, named observations ${y}$ of Gaussian process, are denoted  by the variable
\begin{equation}
    {y} = f(\boldsymbol{x})+{e}, \quad {e} \sim \mathcal{N}(0,\sigma_n^2),
\end{equation}
where ${e}$ is the additive independent identically distributed Gaussian noise $\mathcal{N}$ with variance $\sigma_n^2$.

Assuming there are $n$ training points and $n^*$ test points when making inferences, the prior is
\begin{equation}
  \left[
    \begin{array}{c}
      \boldsymbol{y} \\
      \boldsymbol{f}^*
    \end{array}
    \right] 
  \sim \mathcal{N}
  \left(
    \boldsymbol{0},
  \left[
    \begin{array}{cc}
      K(\boldsymbol{X},\boldsymbol{X})+\sigma^2_n\boldsymbol{I} & K(\boldsymbol{X},\boldsymbol{X}^*) \\
      K(\boldsymbol{X}^*,\boldsymbol{X}) & K(\boldsymbol{X}^*,\boldsymbol{X}^*)
    \end{array}
    \right]
  \right),
\end{equation}
where $(\boldsymbol{X},\boldsymbol{y})$ are the training data with dimension $n\times d$ and $n \times 1$ respectively;
$\boldsymbol{X}^*$ are the test data points with dimension $n^* \times d$ on which the prediction is made;
$\boldsymbol{f}^*$ are the $n^*$ dimension predicted value; $K(\boldsymbol{X},\boldsymbol{X}^*)$ denotes the $n \times n^*$ matrix of the covariances evaluated at all pairs of training and test
points, and similarly for $K(\boldsymbol{X},\boldsymbol{X})$, $K(\boldsymbol{X}^*,\boldsymbol{X}^*)$ and $K(\boldsymbol{X}^*,\boldsymbol{X})$.

The posterior is
\begin{equation}
\boldsymbol{f}^*| \boldsymbol{X}^*,\boldsymbol{X},\boldsymbol{y} \sim \mathcal{N}\left(\mathrm{E}\{\boldsymbol{f}^*\},\mathrm{cov}\{\boldsymbol{f}^*\}\right),
\end{equation}
where
\begin{equation}
    \begin{array}{rl}
      \mathrm{E}\{\boldsymbol{f}^*\} =&\! K(\boldsymbol{X}^*,\boldsymbol{X})\left[K(\boldsymbol{X},\boldsymbol{X})+\sigma^2_n\boldsymbol{I}\right]^{-1}\boldsymbol{y} \qquad \text{and}\\
      \mathrm{cov}\{\boldsymbol{f}^*\} =&\! K(\boldsymbol{X}^*,\boldsymbol{X}^*) - K(\boldsymbol{X}^*,\boldsymbol{X})\left[K(\boldsymbol{X},\boldsymbol{X})+\sigma^2_n\boldsymbol{I}\right]^{-1}K(\boldsymbol{X},\boldsymbol{X}^*).
    \end{array}
\end{equation}

General GP regression is equivalent to Bayesian linear regression with an infinite number of basis functions. The kernel function adopted in this work is the squared exponential kernel (Gaussian kernel). The squared exponential kernel $k_{SE}$ is defined as
\begin{equation} \label{eq:gaussiankernel}
    k_{SE}(\boldsymbol{x},\boldsymbol{x}') = 
    \sigma^2\exp(\frac{-\left \| \boldsymbol{x}-\boldsymbol{x}'\right \|^2}{2l^2}),
\end{equation}
where $l$ is the length scale and $\sigma$ is the signal standard deviation. $\left \| \boldsymbol{x}-\boldsymbol{x}'\right \|$ is the Euclidean norm of $\boldsymbol{x}-\boldsymbol{x}'$, representing the distance between $\boldsymbol{x}$ and $\boldsymbol{x}'$.

The length scale $l$ can be interpreted as a measure of how closely the points $\boldsymbol{x}$ and $\boldsymbol{x}'$ need to be to significantly influence each other. When the length scale $l$ is a scalar, the kernel is considered isotropic. In contrast, if $l$ varies for each dimension of the data points, the kernel is referred to as an automatic relevance determination (ARD) kernel ~\cite{GaussianProcessesforMachineLearning2006}. The ARD kernel allows the model to determine the relevance of each dimension separately, providing a feature selection ability. In this case, the expressions related to the length scale in Eq. ~(\ref{eq:gaussiankernel}) are modified to a summation form. Additionally, the model's flexibility can be extended by adding multiple kernels.

The optimal parameters of the GP are found by maximizing the log marginal likelihood $ \log p$,
\begin{equation} \label{eq:logmarginallikelihood}
    \log p(\boldsymbol{y}|\boldsymbol{X},\boldsymbol{\theta}) = -\frac{1}{2}\boldsymbol{y}^T\boldsymbol{K}_y^{-1}\boldsymbol{y}
    -\frac{1}{2}\log \left \| \boldsymbol{K}_y \right \|
    -\frac{n}{2}\log 2\pi,
\end{equation}
where $\boldsymbol{K}_y = K(\boldsymbol{X},\boldsymbol{X})+\sigma^2_n\boldsymbol{I}$ is the covariance matrix for the noisy targets $\boldsymbol{y}$ and $n$ is the number of training points. $\boldsymbol{\theta}$ are the hyperparameters, including the length scale of each dimension and signal and noise variance \cite{GaussianProcessesforMachineLearning2006}.

\subsection{Gaussian process for fuel cell modeling}
In our previous work \cite{li2021multi}, a detailed physical fuel cell system model was elaborated and built with MATLAB Simulink, which was assumed the closest to the true system dynamics. To model errors in measurement, Gaussian measurement noises were added to output voltage $V_\mathrm{FC}$ and hydrogen pressure $P_\mathrm{H_2}$. 

The Gaussian process model predicts two system states based on three system inputs. These inputs include the control actions, i.e., hydrogen volumetric flow rate $Q_\mathrm{H_2}$, air volumetric flow rate $Q_\mathrm{air}$, and the current $I$. The predicted states are the fuel cell output voltage $V_\mathrm{FC}$ and the hydrogen pressure $P_\mathrm{H_2}$. It is important to note that the current $I$ represents the workload and cannot be manipulated by the controller. The control target is to maintain a fixed value for the fuel cell voltage, while the hydrogen pressure must be kept below a certain limit to ensure safety.

The target of GP modeling is to build a function $\boldsymbol{f}$ that describes the fuel cell dynamics in the time update $\boldsymbol{x}^{k+1}$,
\begin{equation} \label{eq:2outputsGP}
    \boldsymbol{x}^{k+1} = \boldsymbol{f}(\boldsymbol{u}^k,\boldsymbol{x}^k),
\end{equation}
where
\begin{equation}
    %\begin{array}{rcl}
        \boldsymbol{u}^k  =  [Q^k_\mathrm{H_2}\ Q^k_\mathrm{air}\ I^k]^T \qquad \text{and} \qquad
        \boldsymbol{x}^k  =  [V^k_\mathrm{FC}\ P^k_\mathrm{H_2}]^T,
    %\end{array}
\end{equation}

in which the variable  $\boldsymbol{x}^{k+1}$ is the system states at time step $k+1$. $\boldsymbol{u}^k$ and $\boldsymbol{x}^k$ are the model inputs and measured system states at time step $k$, respectively.

The Gaussian process, however, doesn't support multi-dimensional regression natively. One simple solution is to assume the multi-dimensional states are independent of each other and build a Gaussian process for each state with the same inputs. Eq.~(\ref{eq:2outputsGP}) then becomes two Gaussian processes $f_V$ and $f_P$,

\begin{equation} \label{eq:fvfp}
    %\begin{array}{rcl}
        V^{k+1}_\mathrm{FC}  =  f_V(\boldsymbol{u}^k,V^k_\mathrm{FC}) \qquad \text{and} \qquad 
        P^{k+1}_\mathrm{H_2}  =  f_P(\boldsymbol{u}^k,P^k_\mathrm{H_2}).
    %\end{array}
\end{equation}

To collect the training data points for the Gaussian process, Latin hypercube sampling (LHS) was applied to the inputs $Q_\mathrm{H_2}$, $Q_\mathrm{air}$ and the current $I$. The interval between time steps was 0.5 s. In total, 1000 training data points were collected.

The kernel type used for the two Gaussian processes was a combination of an isotropic squared exponential kernel (Gaussian kernel) and an automatic relevance determination (ARD) squared exponential kernel. The parameters of the Gaussian process model were selected by maximizing the log-likelihood function described in Eq.~(\ref{eq:logmarginallikelihood}). Solving this optimization problem is typically challenging due to its nonlinearity and nonconvexity. In this study, the limited-memory Broyden–Fletcher–Goldfarb–Shanno (L-BFGS) algorithm~\cite{liu1989limited} was employed to efficiently solve the problem.

\begin{figure}[!bhtp]
    \centering
    \includegraphics[width=0.8\linewidth]{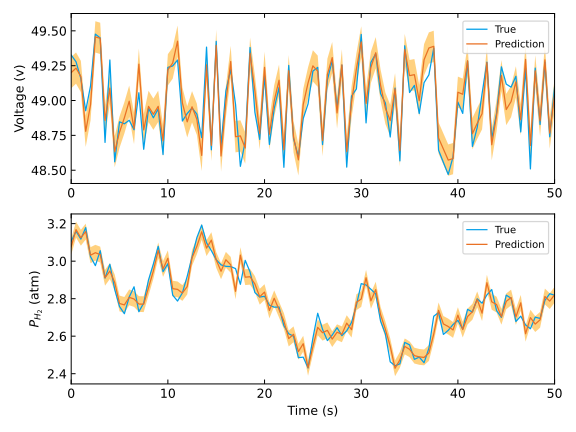}
    \caption{Gaussian process training results. The orange shaded area is the one $\sigma$ confidence interval, the prediction value plus/minus one prediction standard deviation (one $\sigma$).}
    \label{fig:GPprediction}
\end{figure}

After training, the GP model was validated using testing data points. These testing data points were obtained by applying a different set of Latin Hypercube Sampling (LHS) inputs to the fuel cell Simulink model. A total of 300 test points were collected. The prediction results are depicted in Fig. ~\ref{fig:GPprediction}, where the blue line represents the true values and the orange line represents the one-step predictions made by the GP model. The orange shaded area represents the one $\sigma$ confidence interval, which is the prediction value plus/minus one prediction standard deviation (one $\sigma$).

It is evident from the results that the GP model accurately predicted the behavior of the fuel cell in terms of $V_\mathrm{FC}$ and $P_\mathrm{H_2}$. The variance calculated by the GP model served as an indicator of the prediction confidence. The width of the prediction band was wider in regions where the test points were far from the training points, indicating lower prediction accuracy by the GP model. However, in the majority of cases, the prediction band was able to encompass the true value.

\section{Gaussian process MPC design}
\label{GP-MPC design}
The state-space model for MPC is obtained by linearizing the Gaussian process models. The linearized or Gaussian process $f_V$  shown in Eq.~(\ref{eq:fvfp}) then becomes
\begin{equation}
\mathrm{d}V_\mathrm{FC} = \begin{bmatrix} \dfrac{\partial f_V}{\partial Q_\mathrm{H_2}} & \dfrac{\partial f_V}{\partial Q_\mathrm{air}} & \dfrac{\partial f_V}{\partial I}  \end{bmatrix} \! \begin{bmatrix} \mathrm{d}Q_\mathrm{H_2}\\ \mathrm{d}Q_\mathrm{air} \\ \mathrm{d}I \end{bmatrix},
\end{equation}
where the partial derivative is taken on the latest system states which will update each time step. The partial derivative of the GP with Gaussian kernel can be solved explicitly, but the forward mode automatic differentiation (AD) method \cite{baydin2018automatic} is used here for simplicity and flexibility. The same expression is true for the Gaussian process $f_P$.

The discrete-time state-space model of the fuel cell used for control is written as
\begin{equation} \label{eq:statespacemodelgpfc}
  %\begin{array}{rl}
    \boldsymbol{x}^{k+1}\!= \boldsymbol{A} \boldsymbol{x}^k + \boldsymbol{B} \boldsymbol{u}^k
   \qquad \text{and} \qquad 
    \boldsymbol{y}^k\!= \boldsymbol{C} \boldsymbol{x}^k,
  %\end{array}
\end{equation}
where the state vector $\boldsymbol{x}^k$ at sample index $k$ is
\begin{equation} \label{eq:ssmxgpfc}
  \boldsymbol{x}^k = \begin{bmatrix} V_\mathrm{FC}^{k}\\ P_\mathrm{H_2}^{k}\\ \mathrm{d}I^{k}\\ Q_\mathrm{H_2}^{k}\\ Q_\mathrm{air}^{k} \end{bmatrix} \qquad \text{with input} \qquad  \boldsymbol{u}^k = \begin{bmatrix} \mathrm{d}Q_\mathrm{H_2}^k\\ \mathrm{d}Q_\mathrm{air}^k  \end{bmatrix} \qquad \text{and output} \qquad \boldsymbol{y}^k = \begin{bmatrix} V_\mathrm{FC}^{k} \end{bmatrix},
\end{equation}

and state-space matrices are
\begin{equation} \label{eq:ssmABCgpfc}
  %\begin{array} {ll}
    \boldsymbol{A}  = \begin{bmatrix}
      1\ & 0\ & \frac{\partial f_V}{\partial I}\ & 0\ & 0\\
      0\ & 1\ & \frac{\partial f_P}{\partial I}\ & 0\ & 0\\
      0\ & 0\ & 0\ & 0\ & 0\\
      0\ & 0\ & 0\ & 1\ & 0\\
      0\ & 0\ & 0\ & 0\ & 1\\
      \end{bmatrix},
    \boldsymbol{B}  = \begin{bmatrix}
      \frac{\partial f_V}{\partial Q_\mathrm{H_2}}\ & \frac{\partial f_V}{\partial Q_\mathrm{air}} \\
      \frac{\partial f_P}{\partial Q_\mathrm{H_2}}\ & \frac{\partial f_P}{\partial Q_\mathrm{air}} \\
      0\ & 0\\
      1\ & 0\\
      0\ & 1
      \end{bmatrix} , \qquad \text{and} \qquad 
    \boldsymbol{C}  = \begin{bmatrix} 1\ 0\ 0\ 0\ 0\ \end{bmatrix}.
  %\end{array}
\end{equation}

The actuator increments were selected as the system inputs.
Consequently, $Q_\mathrm{H_2}^{k}$ and $Q_\mathrm{air}^{k}$ were added into the state vector to help impose appropriate constraints.

A Quadratic Programming (QP) problem will be solved at each time step to obtain {the optimal} control inputs in this MPC problem formulation
\begin{equation} \label{eq:MPCcostfunctiongpfc}
  \min_{\boldsymbol{u}^0,\boldsymbol{u}^1,...,\boldsymbol{u}^{H_u-1}} \boldsymbol{J}(\boldsymbol{u}^k) = 
    \sum_{k=1}^{H_{p}} \left \| \boldsymbol{y}^k - \boldsymbol{r} \right \|_{\boldsymbol{Q}}^2
        +\sum_{k=0}^{H_{u}-1} \left \| \boldsymbol{u}^k \right \|_{\boldsymbol{R}}^2
        + \rho \sum_{k=1}^{H_{p}} \left \| \boldsymbol{\epsilon}^k \right \|^2,
\end{equation}
{where $H_p$ and $H_u$ are predictions and control horizons.}
Superscript $k$ represents the time step, $k = 0, 1, \dots, H_p$, and $k=0$ refers to the current time step. $\boldsymbol{r}$ denotes the control reference.
$\boldsymbol{Q}$ and $\boldsymbol{R}$ are weight-tuning parameters for reference tracking and control inputs, respectively. The discrete-time state-space mode can be further written as 
\begin{equation} \label{eq:MPCconstraintsgpfc}
  \begin{array}{rl}
    \boldsymbol{x}^{k+1} &\!= \boldsymbol{A^d} \boldsymbol{x}^k + \boldsymbol{B^d} \boldsymbol{u}^k\\
    \boldsymbol{y}^k &\!= \boldsymbol{C^d} \boldsymbol{x}^k\\
    \end{array},
\end{equation}
in which $\boldsymbol{A^d}$, $\boldsymbol{B^d}$, and $\boldsymbol{C^d}$ are state-space matrices $\boldsymbol{A}$, $\boldsymbol{B}$ and $\boldsymbol{C}$ in discrete-time. The constraints of the equation \ref{eq:MPCconstraintsgpfc} can be expressed as
\begin{equation}
  \begin{array}{rl}
    \boldsymbol{u}_\mathrm{lb} &\!\leq \boldsymbol{u}^k \leq \boldsymbol{u}_\mathrm{ub}\\
    \mathrm{d}\boldsymbol{u}_\mathrm{lb} &\!\leq \boldsymbol{u}^{k} - \boldsymbol{u}^{k-1} \leq \mathrm{d}\boldsymbol{u}_\mathrm{ub}\\
    \boldsymbol{x}_\mathrm{lb} &\!\leq \boldsymbol{x}^k \leq \boldsymbol{x}_\mathrm{ub} + \boldsymbol{\epsilon}^k\\
  \end{array},
\end{equation}
where $\boldsymbol{u}_\mathrm{lb}$, $\boldsymbol{u}_\mathrm{ub}$, $\boldsymbol{x}_\mathrm{lb}$, and $\boldsymbol{x}_\mathrm{ub}$ are the lower bounds and upper bounds of inputs $\boldsymbol{u}$ and states $\boldsymbol{x}$, respectively. $\mathrm{d}\boldsymbol{u}_\mathrm{lb}$ and $\mathrm{d}\boldsymbol{u}_\mathrm{ub}$ are the lower bounds and upper bounds of inputs change rate. $\boldsymbol{\epsilon}$ is the slack variable introduced to soft constraints and $\boldsymbol{\epsilon}^k \geq \boldsymbol{0}$. Further constraints are
\begin{equation}
  \begin{array}{rl}
    P_\mathrm{H_2}^k &\!\leq P_\mathrm{H_2}^{limit} + \epsilon^k \\
    P_\mathrm{H_2}^k &\!\leq P_\mathrm{H_2}^{limit} + \epsilon^k - \alpha \sigma_p \Delta P_\mathrm{H_2} 
  \end{array},
\end{equation}
% some detailed symbols
where $P_\mathrm{H_2}^{limit}$ is the upper bound of $P_\mathrm{H_2}$ and $\rho$ is a nonnegative scalar to control the magnitude of penalizing soft constraint violations. In addition, $\boldsymbol{u}^{-1} = \boldsymbol{u}_\mathrm{init}$ which is the latest applied control input, and $\boldsymbol{x}^0 = \boldsymbol{x}_\mathrm{init}$ which is the latest measured value, i.e., the state feedback.

The additional inequality for $P_\mathrm{H_2}$ as a compensation for model errors is
\begin{equation} \label{eq:PH2additionalconstraint}
    P_\mathrm{H_2}^k  \leq P_\mathrm{H_2}^{limit} + \epsilon^k - \alpha \sigma_p \Delta P_\mathrm{H_2},
\end{equation}
% where $P_\mathrm{H_2}^{limit}$ is the upper bound of $P_\mathrm{H_2}$;
where $\sigma_p$ is the prediction variance when making an inference with Gaussian process $f_P$ on the data point in the current time step. 
$\Delta P_\mathrm{H_2}$ is the possible $P_\mathrm{H_2}$ move with respect to the possible calculated inputs
\begin{equation} \label{eq:DeltaP_H2}
    \Delta P_\mathrm{H_2} = \sum_{i=0}^{k-1}(\mathrm{d}Q_\mathrm{H_2}^i \frac{\partial f_P}{\partial Q_\mathrm{H_2}}  + \mathrm{d}Q_\mathrm{air}^i \frac{\partial f_P}{\partial Q_\mathrm{air}}),
\end{equation}
where $\alpha$ is a tuning parameter. This term, i.e., $\alpha \sigma_p \Delta P_\mathrm{H_2}$, can be understood from an intuitive point of view.
Typically, a bigger possible move will lead to a bigger uncertainty. Thus, a redundancy proportional to the possible movement size and prediction uncertainty in the constraint is desirable to compensate for the model imperfections. Since the $P_\mathrm{H_2}$ limit is from above, the added constraint in Eq.~(\ref{eq:PH2additionalconstraint}) is only meaningful when the possible move is positive.

To be precise, this constraint on $P_\mathrm{H_2}^k$ is simplified from the stochastic MPC constraint formation.
In stochastic MPC, a chance constraint is
\begin{equation} \label{eq:chanceconstraint}
    p(\boldsymbol{x}^k \leq \boldsymbol{x}_\mathrm{ub}) \geq \eta,
\end{equation}
where $\eta$ denotes the confidence level. When $\eta$ is 0.95, it is 
\begin{equation} \label{eq:chanceconstraintsigma}
    \boldsymbol{x}^k \leq \boldsymbol{x}_\mathrm{ub} - 2\Sigma^k,
\end{equation}
where $\Sigma^k$ is the covariance matrix of $\boldsymbol{x}^k$. Here $\boldsymbol{x}^k$ actually represents its mean value.
The $\Sigma^k$ is estimated from the uncertainty propagation.
The specific form for $P_\mathrm{H_2}$ of Eq.~(\ref{eq:chanceconstraintsigma}) is:
\begin{equation}
    P_\mathrm{H_2}^k \leq P_\mathrm{H_2}^{limit} - 2 \sigma_p^k,
\end{equation}

For every step in the prediction horizon,
the $P_\mathrm{H_2}^k$ variance is propagated along with the uncertainty in $\boldsymbol{B}$ matrix.
The linearized form:
\begin{equation}
    P_\mathrm{H_2}^{k+1} = P_\mathrm{H_2}^k + \mathrm{d}Q_\mathrm{H_2}^k \frac{\partial f_P}{\partial Q_\mathrm{H_2}}  + \mathrm{d}Q_\mathrm{air}^k \frac{\partial f_P}{\partial Q_\mathrm{air}}.
\end{equation}

The variance estimate update is
\begin{equation}
    (\sigma_p^{k+1})^2 = (\sigma_p^{k})^2 + (\mathrm{d}Q_\mathrm{H_2}^k)^2 \mathrm{var}(\frac{\partial f_P}{\partial Q_\mathrm{H_2}}) + (\mathrm{d}Q_\mathrm{air}^k)^2 \mathrm{var}(\frac{\partial f_P}{\partial Q_\mathrm{air}}).
\end{equation}

The variance of the partial derivative taken at the initial step is approximated by
\begin{equation}
\begin{array}{rl}
    \mathrm{var}(\dfrac{\partial f_P}{\partial Q_\mathrm{H_2}})&\!= \mathrm{var}(\dfrac{f_P(Q_\mathrm{H_2}+\delta Q_\mathrm{H_2}) - f_P(Q_\mathrm{H_2})}{\delta Q_\mathrm{H_2}})\\
    & = \dfrac{2}{(\delta Q_\mathrm{H_2})^2} (\sigma_p)^2 \\ [3mm]
    & = \alpha_1 \mathrm{E}^2\{\dfrac{\partial f_P}{\partial Q_\mathrm{H_2}}\} (\sigma_p)^2 
\end{array},
\end{equation}
where ${2}/{(\delta Q_\mathrm{H_2})^2}$ is assumed to be proportional to $\mathrm{E}^2\{{\partial f_P}/{\partial Q_\mathrm{H_2}}\}$ value with a constant value $\alpha_1$. The $\mathrm{E}\{{\partial f_P}/{\partial Q_\mathrm{H_2}}\}$ is actually the ${\partial f_P}/{\partial Q_\mathrm{H_2}}$ value calculated at the initial time step.
Thus, at each time step in the prediction horizon, the variance increases with a magnitude of
\begin{equation}
    \alpha_1 (\sigma_p)^2 (\mathrm{d}Q_\mathrm{H_2}^k)^2 \mathrm{E}^2\{\frac{\partial f_P}{\partial Q_\mathrm{H_2}}\} + \alpha_2 (\sigma_p)^2 (\mathrm{d}Q_\mathrm{air}^k)^2 \mathrm{E}^2\{\frac{\partial f_P}{\partial Q_\mathrm{air}}\}.
\end{equation}

Summing over the steps until time step $k$, and viewing all constant coefficient as the same, gives the total standard error of time step $k$:
\begin{equation} \label{eq:sumoverallstepssquareform}
   \sigma_p^k = \alpha \sigma_p \sum_{i=0}^{k-1} \sqrt{ (\mathrm{d}Q_\mathrm{H_2}^k)^2 \mathrm{E}^2\{\frac{\partial f_P}{\partial Q_\mathrm{H_2}}\} + (\mathrm{d}Q_\mathrm{air}^k)^2 \mathrm{E}^2\{\frac{\partial f_P}{\partial Q_\mathrm{air}}\}}.
\end{equation}
All constants are written together as $\alpha$.
However this form is too conservative for constraint handling and requires high computational effort. The root part is approximated as
\begin{equation}
    \mathrm{d}Q_\mathrm{H_2}^k \mathrm{E}\{\frac{\partial f_P}{\partial Q_\mathrm{H_2}}\} + \mathrm{d}Q_\mathrm{air}^k \mathrm{E}\{\frac{\partial f_P}{\partial Q_\mathrm{air}}\}.
\end{equation}

Then the Eq.~(\ref{eq:sumoverallstepssquareform}) becomes 
\begin{equation}
    \sigma_p^k = \alpha \sigma_p \Delta P_\mathrm{H_2},
\end{equation}
where $\Delta P_\mathrm{H_2}$ is shown in the Eq.(\ref{eq:DeltaP_H2}). This also gives an intuitive interpretation of the constraint redundancy.
Meanwhile, the constant 2 and $\alpha$ are integrated into the coefficient $\alpha$ in the constraints described by Eq.~(\ref{eq:MPCconstraintsgpfc}). In the present study, this does not necessarily give a 95\% confidence level, and $\alpha$ should be tuned according to the performance and safety requirements.

\section{Simulation set-up}\label{Experimental set-up}
The simulation of the controller performance was conducted on the Simulink model detailed in \cite{li2021multi}. This model is viewed as the true system dynamics. Gaussian measurement noises were added to voltage $V_\mathrm{FC}$ and hydrogen pressure $P_\mathrm{H_2}$.

\begin{figure}[htp]
    \centering
    \includegraphics[width=\linewidth]{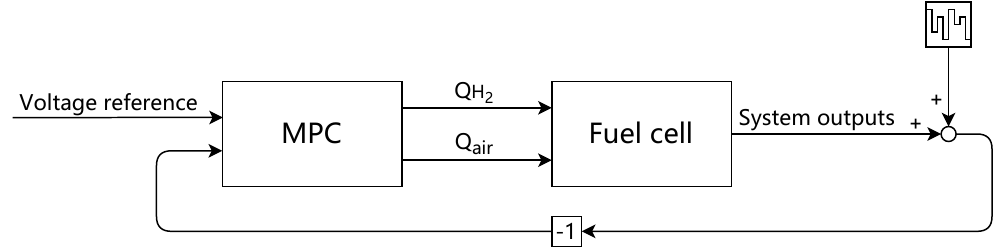}
    \caption{MPC control process}
    \label{fig:fuelcellcontroldiagram_MPC}
\end{figure}

Figure~\ref{fig:fuelcellcontroldiagram_MPC} gives the illustration of the MPC control process.

The formulation of MPC is shown in \cite{li2021multi}. Besides, the constraints regarding the input change rate and hydrogen pressure were added.

\begin{figure}[htp]
    \centering
    % \fontsize{6}{8}\selectfont
    % \includegraphics[width=\linewidth]{2_modeling/ICEgassystemdiagram.pdf}
    \includegraphics[width=\linewidth]{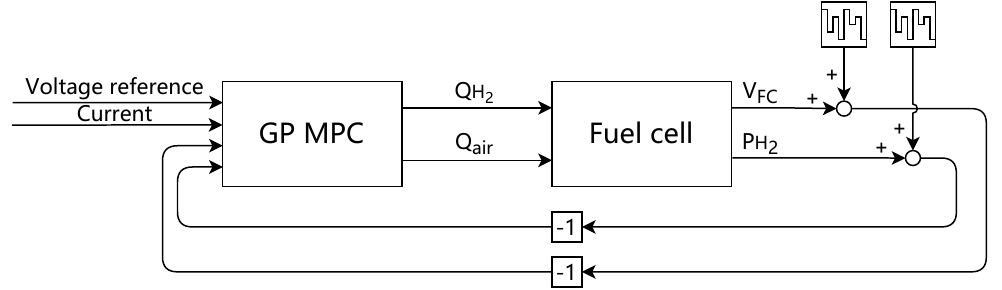}
    \caption{GP-MPC control process}
    \label{fig:fuelcellcontroldiagram_GPMPC}
\end{figure}

The illustration of the GP-MPC control process is shown in Fig.~\ref{fig:fuelcellcontroldiagram_GPMPC}.

Compared with the MPC controller shown in Fig.~\ref{fig:fuelcellcontroldiagram_MPC}, the GP-MPC only needs the system information of $V_\mathrm{FC}$ and $P_\mathrm{H_2}$, whereas MPC requires other information such as oxygen and nitrogen pressure. This is a desirable aspect in practice.

In the fuel cell problem, the state constraint is the limitation of the hydrogen pressure $P_\mathrm{H_2}$. The hydrogen pressure in the pipe should be under a certain value to ensure safety, which is 2.5 $atm$ here. Only slack variables for state upper bounds are introduced. The input $Q_\mathrm{H_2}$ is limited between 100 and 400 $lpm$ (liters per minute) and $Q_\mathrm{air}$ is limited within 300 to 700 $lpm$. The change rate of the inputs is constrained within -40 to 20 lpm. The fuel cell system size is 6 kW.

The time step for the MPC controller is 0.5 s. The QP problem is solved at each step, then the solved $Q_\mathrm{H_2}$ + $\mathrm{d} Q_\mathrm{H_2}$ and $Q_\mathrm{air}$ + $\mathrm{d} Q_\mathrm{air}$ are applied to the fuel cell plant.

\section{Simulation results}
\label{Simulation results}

The performance of GP-MPC and MPC with physical models explained in \cite{li2021multi} were compared.
Two test scenarios were chosen, one was the typical step disturbance applied on the working load, the current; the other was a mixture of slope and step working load changes.

\begin{figure}[!bhtp]
    \centering
    % \fontsize{6}{8}\selectfont
    % \includegraphics[width=\linewidth]{2_modeling/ICEgassystemdiagram.pdf}
    \includegraphics[width=0.8\linewidth]{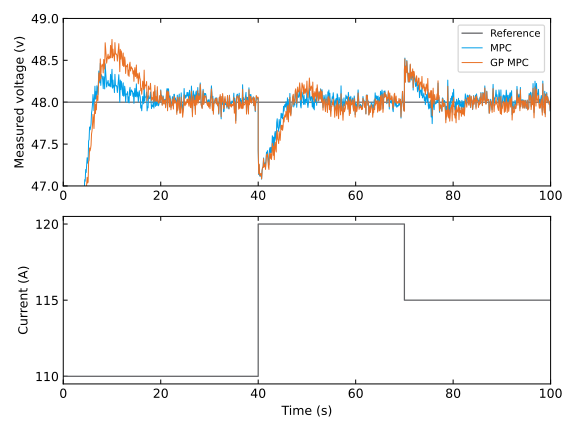}
    \caption{Output voltage under the current step disturbance.}
    \label{fig:MPCvsGPMPC1}
\end{figure}

\begin{figure}[!bhtp]
    \centering
    % \fontsize{6}{8}\selectfont
    % \includegraphics[width=\linewidth]{2_modeling/ICEgassystemdiagram.pdf}
    \includegraphics[width=0.8\linewidth]{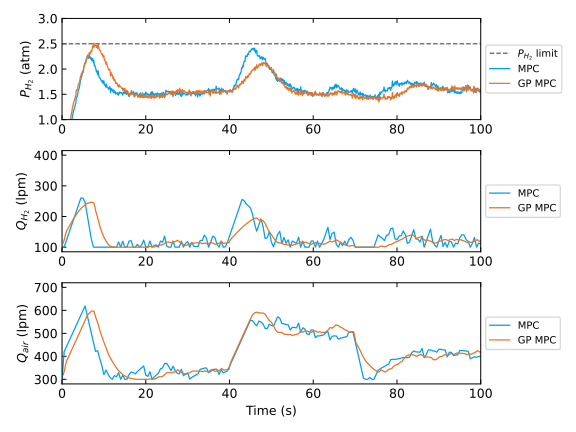}
    \caption{Constraint handling and system inputs of MPC and GP-MPC.}
    \label{fig:MPCvsGPMPC2}
\end{figure}

Figure~\ref{fig:MPCvsGPMPC1} shows the GP-MPC voltage tracking performance compared with MPC under step workload disturbance, and Fig.~\ref{fig:MPCvsGPMPC2} presents the corresponding $P_\mathrm{H_2}$ behavior and system inputs. At the beginning of the experiment, MPC and GP-MPC had similar rise-up traces, which were limited by the input change rate constraint. MPC had a lower overshoot than GP-MPC benefiting from the accurate system model. The overshoot for GP-MPC was 0.60 V and for MPC was 0.42 V. Both controllers can successfully satisfy the $P_\mathrm{H_2}$ safety requirements. When the current increased suddenly, MPC and GP-MPC drove the voltage back to the reference at a similar pace, but MPC responded faster. To arrive at 47.9 V again, MPC took 5.2 s and GP-MPC took 6.5 s. It can be clearly seen that the inputs increased following a straight line, indicating the activation of the change rate limits. However, similar to the start of the system, the GP-MPC calculated $Q_\mathrm{H_2}$ increment was rather conservative in comparison with MPC. This is because the added constraint in Eq.~(\ref{eq:PH2additionalconstraint}) enforced the controller to take a more cautious action when increasing $Q_\mathrm{H_2}$ which directly increased the $P_\mathrm{H_2}$. In contrast, $Q_\mathrm{air}$ was adjusted aggressively. When the current suddenly dropped at time 75 s, MPC and GP-MPC had a similar performance in pulling back the voltage. The steady-state tracking behavior of the two controllers was comparable and the safety constraint was satisfied throughout the process.

% escalation rate

% figure reference format; inside text: Fig.

% also equation table etc. format

\begin{figure}[htp]
    \centering
    % \fontsize{6}{8}\selectfont
    % \includegraphics[width=\linewidth]{2_modeling/ICEgassystemdiagram.pdf}
    \includegraphics[width=0.8\linewidth]{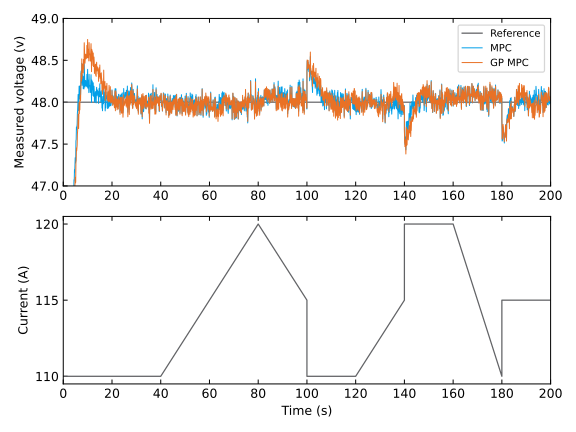}
    \caption{Output voltage under the current step and slope disturbance.}
    \label{fig:MPCvsGPMPC_rampI1}
\end{figure}

\begin{figure}
     \centering
    % \fontsize{6}{8}\selectfont
    % \includegraphics[width=\linewidth]{2_modeling/ICEgassystemdiagram.pdf}
    \includegraphics[width=0.8\linewidth]{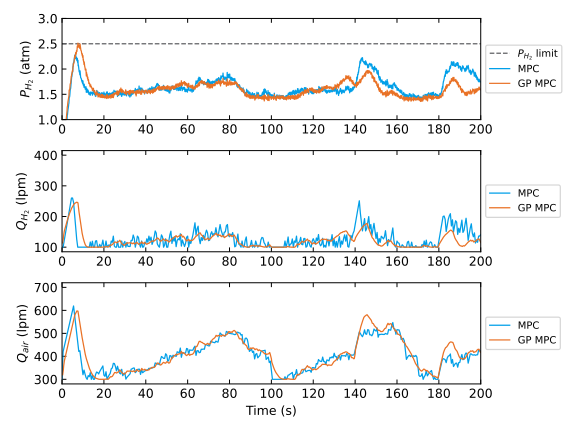}
    \caption{Constraint handling and system inputs of MPC and GP-MPC.}
    \label{fig:MPCvsGPMPC_rampI2}
\end{figure}

Figures~\ref{fig:MPCvsGPMPC_rampI1} and ~\ref{fig:MPCvsGPMPC_rampI2} exhibit the GP-MPC and MPC behavior under slope and step current changes. The behavior was similar to the step change scenario, and GP-MPC and MPC had equivalent performance most of the time. The $P_\mathrm{H_2}$ constraint was satisfied in all cases.

\section{Discussion}
\label{Discussion}
In this paper, a novel GP-MPC was proposed for the fuel cell control task. Its performance was validated on the Simulink model. Although the Gaussian process had been used to build fuel cell related models \cite{He2016GP, XIE202030942}, it was the first time application of GP-MPC on fuel cell control to the author's knowledge. 
Even though the MPC based on the physical model had a better performance in terms of the overshoot during system start-up, the GP-MPC required less system information as compared to the work in \cite{li2021multi}, which needed fewer sensors and was more economical. The modeling process of GP-MPC relied on collected data, whereas physics-based MPC required a deeper understanding of the system inner principles.
The GP-MPC considered the safety requirements regarding pressure and actuators as well, which was missed in \cite{GRUBER2012205} and \cite{ouyang2017nonlinear}. Though based on nonlinear GP models, the MPC framework took linearized models and reduced the computational burden, whereas the nonlinear MPC in \cite{ZIOGOU2018656} made it difficult to implement and solve. To compensate for the model imperfection, one constraint taking prediction variance and possible moves into account was added. This gave a satisfactory constraint handling result under model error and linearization error.

\section{Conclusion}

In summary, a Gaussian process MPC was developed to control the fuel cell voltage. Two Gaussian processes were used to predict the voltage and hydrogen pressure. 

The state-space models were formed based on the linearized Gaussian process. A special inequality utilizing GP prediction variance was added to compensate for the model imperfections in satisfying constraints. The simulation results showed the GP-MPC kept the voltage at the desired 48 V while satisfying the safety constraints including the hydrogen pressure limit of 2.5 atm and input change rate limit all the time under a workload disturbance ranging from 110-120 A. The constraint handling of GP-MPC gave a conservative action, but the safety requirements were satisfied well benefiting from this characteristic.
As compared to the MPC with the knowledge of the detailed underlying system dynamics, the GP-MPC has a 43\% higher overshoot and 25\% slower response time, but the GP-MPC eliminates the requirement of the underlying true system model and the model simplification process involving expert knowledge.

\section*{Acknowledgements}
The first author would like to acknowledge the Competence Centre Combustion Processes, KCFP, and the Swedish Energy Agency (grant number 22485-4) for the financial support.
The Chinese Scholarship Council is also thanked for the sponsorship of living expenses during the first author’s research.
Rolf Johansson is a member of the LCCC Linnaeus Center and the eLLIIT Excellence Center at Lund University.

%Bibliography
\bibliographystyle{IEEEtran}
\bibliography{mybib}
\end{document}